# A Decision-Aided Parallel SC-List Decoder for Polar Codes

Bin Li, Hui Shen, Kai Chen

*Abstract*—In this paper, we propose a decision-aided scheme for parallel SC-List decoding of polar codes. At the parallel SC-List decoder, each survival path is extended based on multiple information bits, therefore the number of split paths becomes very large and the sorting to find the top *L* paths becomes very complex. We propose a decision-aided scheme to reduce the number of split paths and thus reduce the sorting complexity.

## I. INTRODUCTION

Polar codes are a major breakthrough in coding theory [1]. They can achieve Shannon capacity with a simple encoder and a simple successive cancellation decoder, both with low complexity of the order of $O(N \log N)$, where $N$ is the code block size. When the code block size is long enough, the simple SC decoder can approaches Shannon capacity. But for short and moderate lengths, the error rate performance of polar codes with the SC decoding is not as good as LDPC or turbo codes. A new SC-list decoding algorithm was proposed for polar codes recently [2], which performs better than the simple SC decoder and performs almost the same as the optimal ML (maximum likelihood) decoding at high SNR. In order to improve the low minimum distance of polar codes, the concatenation of polar codes with simple CRC was proposed [2], and it was shown that a simple concatenation scheme of polar code (2048, 1024) with a 16-bit CRC using the SC-List decoding can outperform Turbo and LDPC codes [3] and is 0.25dB from theoretical limit using adaptive SC-List decoder with very large list size [4].

Although Polar codes can provide good error-rate performance, the SC-List decoder works in a serial fashion. They decode information bits one-by-one. It is very hard to achieve a low decoding latency due to this serial decoding. Parallel SC-List decoder [5][6] were proposed to reduce the decoding latency. Since the parallel SC-List decoder splits each survival path into many paths based on multiple information bits, the number of all split paths becomes very large and the sorting of these split paths becomes very complex. In this paper, we propose a decision-aided scheme to reduce to the number of split paths and therefore the sorting complexity can be reduced. The idea is initially presented in [7], and this paper further extends the idea into parallel SC-List decoding.

In section II, we propose a decision-aided scheme for parallel SC-List decoder for polar codes, and in section III we provide complexity-reduction analysis and some performance simulation results. Finally we draw some conclusions.

## II. A DECISION-AIDED PARALLEL SC-LIST DECODER

### A. Polar Codes and Parallel SC-List Decoder

Let $F = \begin{bmatrix} 1 & 0 \\ 1 & 1 \end{bmatrix}$, $F^{\otimes n}$ is a $N \times N$ matrix, where $N = 2^n$, $\otimes n$ denotes $n$th Kronecker power, and $F^{\otimes n} = F \otimes F^{\otimes(n-1)}$. Let the $n$-bit binary representation of integer $i$ be $b_{n-1}, b_{n-2}, ..., b_0$. The $n$-bit representation $b_0, b_1, ..., b_n$ is a bit-reversal order of $i$. The generator matrix of polar code is defined as $G_N = B_N F^{\otimes n}$, where $B_N$ is a bit-reversal permutation matrix. The polar code is generated by

$$x_1^N = u_1^N G_N = u_1^N B_N F^{\otimes n} \qquad (1)$$

where $x_1^N = (x_1, x_2, ..., x_N)$ is the encoded bit sequence, and $u_1^N = (u_1, u_2, ..., u_N)$ is the encoding bit sequence. The bit indexes of $u_1^N$ are divided into two subsets: the one containing the information bits and the other containing the frozen bits. For simplicity, the frozen bits are set "0"s.

After transmitting $x_1^N$ through a channel, we have the received signal $y_1^N$. The conventional SC-List decoder keeps tracking $L$ survival paths and extends each survival path into two paths according to current decoding bit $u_k = 0$ and $u_k = 1$. After splitting all of $L$ survival paths, $2L$ split paths are obtained but only top $L$ paths with larger path metrics are maintained.

Rather than extending the paths bit by bit, the parallel SC-List decoder $u_1^N$ also keeps tracking $L$ survival paths but extends each survival path into $2^m$ paths according to current decoding bits $(u_{km+1}, u_{km+2}, \cdots, u_{km+m})$, as shown in Fig. 1. After splitting all of $L$ survival paths, $2^m \times L$ split paths are obtained. In order to maintain the top $L$ paths with larger path metrics, we need to sort and find the $L$ top paths among these $2^m \times L$ split paths. This sorting operation grows very complex when $m$ is large.

Bin Li, Hui Shen and Kai Chen are with the Communications Technology Research Lab., Huawei Technologies, Shenzhen, P. R. China (e-mail:{binli.binli, henry.shen, chenkai.chris}@huawei.com).

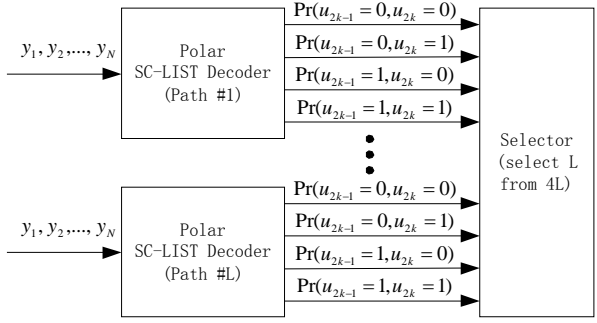

Fig. 1. Parallel SC-List Decoder, $m=2$.

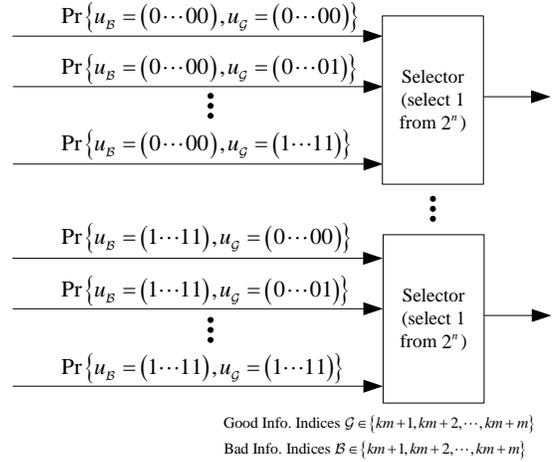

Good Info. Indices $\mathcal{G} \in \{km+1, km+2, \cdots, km+m\}$
Bad Info. Indices $\mathcal{B} \in \{km+1, km+2, \cdots, km+m\}$

Fig. 3. Decision on n good bits.

### B. Decision-Aided SC-List Decoder

According to the principle of polar codes, all information bits have different reliabilities. We divide information bits into two sets according to their reliabilities. A "good" set contains high reliable information bits and a "bad" set contains low reliable information bits. Instead of generating all of $2^m$ split paths for each survival path, we make decision on good bits and only generate a part of these $2^m$ split paths. Suppose that there are $n$ good bits in $(u_{km+1}, u_{km+2}, \cdots, u_{km+m})$ (no frozen bit among them), then we only need to generate $2^{m-n}$ split paths for each survival path and the total number of split paths for sorting is $2^{m-n} \times L$. The decision is made as follows: For any given bit pattern of bad bits, select the path with largest probability among all of $2^n$ bit patterns of good bits. This means that path is split only at the bad bits. Suppose that $u_{km+m}$ is a good bit and the rest are bad bits, for any given bit pattern of $(u_{km+1}, u_{km+2}, \cdots, u_{km+m-1})$, the path with larger probability is selected between the path $(u_{km+1}, u_{km+2}, \cdots, u_{km+m} = 0)$ and the path $(u_{km+1}, u_{km+2}, \cdots, u_{km+m} = 1)$ for all possible $(u_{km+1}, u_{km+2}, \cdots, u_{km+m-1})$, as shown in Fig. 2. More generally, let set $\mathcal{G} \subseteq \{km+1, km+2, \cdots, km+m\}$ with size $|\mathcal{G}| = n$ indicate the good information bits among $(u_{km+1}, u_{km+2}, \cdots, u_{km+m})$; and let set $\mathcal{B}$ indicate the bad information bits, $|\mathcal{B}| \leq m - n$. For each possible $u_{\mathcal{B}} \in \{0,1\}^{|\mathcal{B}|}$, only one path is selected among the $2^n$ paths. A demonstrative graph is shown in Fig. 3.

## III. PERFORMANCE SIMULATIONS AND COMPLEXITY REDUCTION

We simulated the error-rate performance of polar code (2048, 1040) concatenated with 16-bit CRC, with coding rate $R=1/2$, as shown in Fig. 4. The channel is AWGN (additive white Gaussian noise) channel. The decoder is the adaptive SC-List decoder [4] with the maximum list size of $L=32$ and $m=4$. Among 1040 encoding bits for polar code, we select 780 bits with the higher reliability as good bits, which is 75% of all information bits. The rest 260 bits are "bad" bits. We can see that the performance with decision-aided SC-List decoder is the same as that with the original parallel SC-List decoder. If 832 (80% of 1040) information bits are selected as good bits, the performance loss with decision-aided SC-List decoder is negligible compared to the original parallel SC-List decoder.

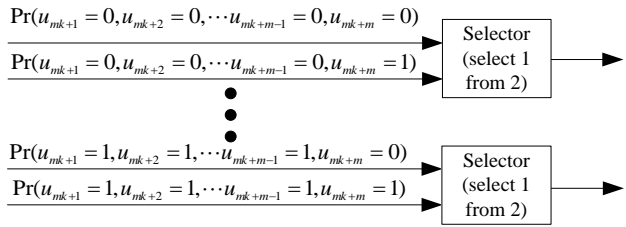

Fig. 2. Decision on one good bit.

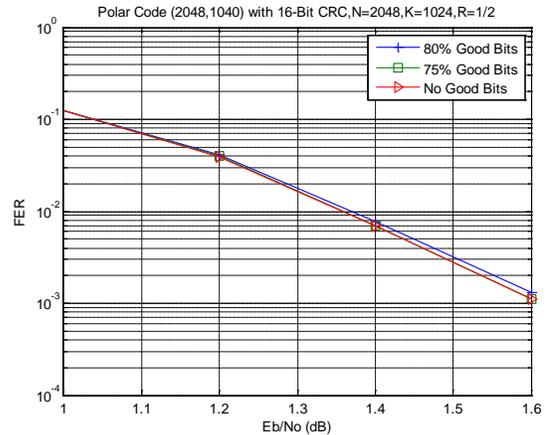

Fig. 4. The FER performance of the polar code (2048, 1040) with 16-bit CRC using the adaptive decision-aided SIC-List decoder, where $m=4$.

TABLE I. PATH EXTENSIONS WITH AND WITHOUT DECISION

| Polar (2048,1040)+16-bit CRC, # Good bits = 780 | | | | |
|---|---|---|---|---|
| Bit Pattern | Good Bit Pattern | $N_1$ | $M_1$ | $M_2$ |
| 0000 | 0000 | 208 | 1 | 1 |
| 0001 | 0000 | 38 | 2 | 2 |
| 0011 | 0000 | 9 | 4 | 4 |
| 0111 | 0000 | 10 | 8 | 8 |
|  | 0001 | 34 | 8 | 4 |
| 1111 | 0001 | 14 | 16 | 8 |
|  | 0011 | 4 | 16 | 4 |
|  | 0111 | 56 | 16 | 2 |
|  | 1111 | 139 | 16 | 1 |

TABLE II. APPEARANCES OF NUMBER OF SPLIT PATHS WI/ WO DECISION

| # of split paths | # of Appearances (wo decision) | # of Appearances (wi decision) |
|---|---|---|
| 2 | 38 | 94 |
| 4 | 9 | 28 |
| 8 | 44 | 24 |
| 16 | 213 | 0 |

The sorting complexity depends on the number of split paths be sorted, and without decision it depends only on the number of information bits in the observed four bits: $(u_{km+1}, u_{km+2}, \cdots, u_{km+m})$, where m=4, but with decision it also depends on the number of good bits in the four observed bits. We analyze all information bit patterns of $(u_{km+1}, u_{km+2}, \cdots, u_{km+m})$ and bit good pattern for each information bit pattern for polar code (2048, 1040) with 780 good bits selected, as shown in Table I, where "0" in the bit pattern represents frozen bit and "1" represents information bit; "1" in good bit pattern represents good bit, $N_1$ represents the number of appearances of a pair of bit pattern and a good bit pattern in one code length. For each pair of bit pattern and a good bit pattern, $M_1$ and $M_2$ represent the numbers of split paths for each survival path without and with decision, respectively. For example, for the pair that the bit pattern is (1111) and good bit pattern is (0001), the number of appearances of this pair in one code length is 14, and each survival path without decision needs to be split into 16 paths but with decision it only needs to be split into 8 paths; For the pair that the bit pattern is (1111) and good bit pattern is (1111), the number of appearances of this pair in one code length is 139, and each survival path without decision needs to be split into 16 paths but it do not need to be split with decision.

Table II shows the number of appearances of different numbers of split paths with and without decision in one code length. For example, there are 44 appearances that a survival path needs to be split into 8 paths without decision but there are only 24 appearances that a survival path needs to be split into 8 paths with decision. If we assume that the sorting complexity is linear to the square of the number of split paths to be sorted, then the complexity without decision is $38 \times 2^2 + 9 \times 4^2 + 44 \times 8^2 + 213 \times 16^2 = 57640$, and the complexity with decision is $94 \times 2^2 + 47 \times 4^2 + 24 \times 8^2 = 2664$, which is 4.6% of that without decision. If we assume that the sorting complexity is linear to the product of the number of split paths with its logarithm function, then the complexity without decision is $38 \times 2 + 9 \times 4 \times 2 + 44 \times 8 \times 3 + 213 \times 16 \times 4 = 14836$, and the complexity with decision is $94 \times 2 + 47 \times 4 \times 2 + 24 \times 8 \times 3 = 1140$, which is 7.7% of that without decision.

Table III shows the number of appearances of all pairs of the bit pattern and good bit pattern for polar code (2048, 1040) with 832 good bits selected. If we assume that the sorting complexity is linear to the square of the number of split paths, then the complexity with decision is $98 \times 2^2 + 52 \times 4^2 + 2 \times 8^2 = 1352$, which is 2.4% of that without decision. If we assume that the sorting complexity is linear to the product of the number of split paths with its logarithm function, then the complexity with decision is $98 \times 2 + 52 \times 4 \times 2 + 2 \times 8 \times 3 = 660$, which is 4.5% of that without decision.

TABLE III. PATH EXTENSIONS WI/WO DECISION

| Polar (2048,1040)+16-bit CRC, # Good bits = 832 | | | | |
|---|---|---|---|---|
| Bit Pattern | Good Bit Pattern | $N_1$ | $M_1$ | $M_2$ |
| 0000 | 0000 | 208 | 1 | 1 |
| 0001 | 0000 | 38 | 2 | 2 |
| 0011 | 0000 | 5 | 4 | 4 |
|  | 0001 | 4 | 4 | 2 |
| 0111 | 0001 | 44 | 8 | 4 |
| 1111 | 0001 | 2 | 16 | 8 |
|  | 0011 | 3 | 16 | 4 |
|  | 0111 | 56 | 16 | 2 |
|  | 1111 | 152 | 16 | 1 |

## IV. CONCLUSION

In this paper, we propose a decision-aided SC-List decoder for parallel implementation architecture. The decision on the split path can significantly reduce to sorting complexity which is very crucial to reduce both complexity and latency of decoder. The simulations show that the proposed scheme does not have performance degradation when the number of good bits is 75% of the total number of information bits.